# Evolution of performance parameters of perovskite solar cells with current-voltage scan frequency

**Enrique H. Balaguera[1,*] and Juan Bisquert[2]**

[1] Escuela Superior de Ciencias Experimentales y Tecnología (ESCET), Universidad Rey Juan Carlos, 28933 Móstoles, Madrid, Spain

[2] Instituto de Tecnología Química (Universitat Politècnica de València-Agencia Estatal Consejo Superior de Investigaciones Científicas), Av. dels Tarongers, 46022 València, Spain

[*]*Corresponding author e-mail:* enrique.hernandez@urjc.es

**Abstract**

Current-voltage measurements are a standard testing protocol to determine the efficiency of any solar cell. However, perovskite solar cells display significant kinetic phenomena that modify the performance at several time scales, due to hysteresis, internal capacitances, and related mechanisms. Here, we develop a method to analyze the current-voltage curves by using large amplitude sinusoids as the excitation waveforms, specifically addressed to determine the influence of cycling frequency on the performance parameters. We solve a system of equations representative of charge collection and recombination, that provide the frequency-dependent dynamical behavior of the internal ion-controlled surface recombination processes that cause open-circuit voltage variations often observed in high performance devices. We analyze several reported experimental data, and we feature the key parameters governing the evolution of hysteresis phenomena as the scan speed is increased in relation to Impedance Spectroscopy.

**Introduction**

In the realm of next-generation photovoltaics, perovskite technology has emerged as a promising platform, captivating researchers from academia and industry worldwide due to the multifaceted advantages.[1–3] Amidst their burgeoning potential, perovskite solar cells harbor, however, omnipresent epiphenomena, such as current-voltage hysteresis[4–7] and related processes[8–11] that make it challenging to accurately assess the true performance of the devices.[12] A careful look at the literature results reveals that, unfortunately, it is not clear so far which dominating effect is responsible for the evolution of the photovoltaic parameters as a function of the voltage scan speed. Therefore, the multicausal nature arising from hysteresis[6,11,12] and the subsequent ups and downs of performance parameters depending on the cycling frequency observed in many later reports now provide us the opportunity to probe perovskite solar cells in different ways to correlate the responses and obtain valuable information about the dominant phenomena.

In this paper, we analyze the manifestation of hysteresis, scanning the voltage by sinusoidal waveforms. This procedure is widely used in a range of ionic/electronic devices such as frequency converters and memristors to determine the dominant kinetic phenomena.[13–16] To our knowledge, the method is not exploited in the study of solar cells. Here we derived new solutions for a perovskite device model[17] to understand the frequency-dependent effects of the persistent hysteresis in perovskite semiconductors by which the different electrical modes (capacitive or inductive) modify the steady-state current responses, inducing in turn memory properties in the devices. We introduce numerical simulation support that explains such dynamic phenomenology, observed in cutting-edge worldwide works, in terms of large transient currents predicted by the physical model and by calculating electrical parameters via current-voltage curves obtained at different conditions. From the reference linear properties of the perovskite devices, we will indeed finalize with general models that produce a robust connection between the dominant kinetic elements estimated by using the small perturbation method of Impedance Spectroscopy and the evolution of photovoltaic parameters. Our work is supported by a series of case studies, showing the effectiveness of the proposed method on real examples, as a crucial prerequisite to fully understand hysteresis problem.

**Results and discussion**

The description of the varied of current-voltage curve results in perovskite solar cells requires the use of an advanced physical theory, such as the dynamical model formulated by the following equations:[18,19]

$$j = C_g \frac{dV}{dt} + J_{rec}(V) + j_d + \frac{dQ_s(v_s)}{dt} - j_{ph} \qquad (1)$$

$$\tau_s \frac{dv_s}{dt} = V - v_s \qquad (2)$$

$$\tau_d \frac{dj_d}{dt} = J_{elect}(V) - j_d \qquad (3)$$

where the external current $j$ is a function of a combination of electronic currents exhibiting very different time scales ($J_{rec}(V)$ and $j_d$), capacitive currents dependent not only on the applied external voltage $V$ (geometric capacitance, $C_g$) but also of a surface ion-controlled potential $v_s$ associated with interface charging processes (given by the term $Q_s(v_s)$ explained below),[20] as well as on the photocurrent $j_{ph}$. Hereafter we use uppercase letters for instantaneous or stationary currents, such as, for instance, the classical recombination term,

$$J_{\text{rec}}(V) = J_{\text{rec}0}e^{\frac{qV}{n_{\text{rec}}k_{\text{B}}T}} \tag{4}$$

where $J_{\text{rec}0}$ is a pre-factor, $q$ represents the electron charge, $k_{\text{B}}$ denotes the Boltzmann's constant, $T$ is the absolute temperature, and $n_{\text{rec}}$ models an ideality factor. Next, we proceed with the justification of the slow kinetic relaxation equations. On the one hand, eq 2 describes the equilibration of $v_{\text{s}}$, in a process that is delayed by ionic dynamics, being $\tau_{\text{s}}$ the relaxation kinetic constant.[21,22] At this point, we introduce $C_{\text{s}}(v_{\text{s}})$ as a characteristic capacitance of photovoltaic perovskites,[23]

$$C_{\text{s}}(v_{\text{s}}) = \frac{dQ_{\text{s}}}{dv_{\text{s}}} = C_{\text{s}0}e^{\frac{qv_{\text{s}}}{n_{\text{s}}k_{\text{B}}T}} \tag{5}$$

which originates from the interaction of ionic and electronic carriers, involving large values in experimental measurements.[24] Here $C_{\text{s}0}$ is a constant pre-factor and $n_{\text{s}}$ models another ideality factor affecting the respective component. On the other hand, the variable $j_{\text{d}}$ with a time-dependent dynamic behavior given by eq 3 represents an additional, delayed recombination current governed by an equilibration time $\tau_{\text{d}}$.[25] This slow electronic current has a steady-state value that also depends exponentially on the voltage $V$; thus, we write

$$J_{\text{elect}}(V) = J_{\text{elect}0}e^{\frac{qV}{n_{\text{d}}k_{\text{B}}T}} \tag{6}$$

Again, $J_{\text{elect}0}$ represents a pre-factor, and $n_{\text{d}}$ an ideality factor, affecting the respective component. In effect, the steady-state current $J(V)$ can be obtained at sufficiently slow sweep velocities,[18,26] when the internal variables $v_{\text{s}}$ and $j_{\text{d}}$ reach their equilibrium values ($v_{\text{s}} \rightarrow V$ and $j_{\text{d}} \rightarrow J_{\text{elect}}(V)$). According to previous mathematical expressions, the partial currents satisfy the following equation in dc conditions:

$$J(V) = J_{\text{rec}}(V) + J_{\text{elect}}(V) \tag{7}$$

If, however, one develops a fast voltage sweep, then the internal surface voltage $v_{\text{s}}$ and/or the electronic current $j_{\text{d}}$ relax but does not reach the quasi-steady-state imposed by the external potential $V$.[27]

The theoretical model is complete and allows the calculation of the results obtained from a variety of experimental methods, such as linear sweep voltammetry (consisting on voltage scan at constant velocity), impedance measurements, and/or transient techniques, as summarized in a recent survey.[17] In this work, we explore a method based on large amplitude sinusoidal sweep scans that is widely used in related areas.[15] The applied perturbation consists of the time dependence,

$$V(t) = V_{\text{p}}\sin(\Omega t + \phi) \tag{8}$$

where $V_{\text{p}}$ is the amplitude voltage of the sweep, $\Omega$ is the angular frequency in units of rad/s ($\Omega = 2\pi f_{\Omega}$ with $f_{\Omega}$ as the frequency) and $\phi$ is the phase that allows one to modulate the initial value $V_0$ of the voltage scanning (at $t = 0$) as

$$\phi = \arcsin\left(\frac{V_0}{V_{\text{p}}}\right) \tag{9}$$

where a negative value indicates a delay ($V_0 < 0$ V) and a positive quantity represents an advance ($V_0 > 0$ V) in time by the amount $\pm \phi/\Omega$ seconds. Note that we consider the sine function in order to analyze a voltage sweep first in forward and then in reverse direction.

The application of the contrary situation (from reverse to forward direction) is explained in the Supporting Information. The analytical expression in time-domain for the surface polarization potential $v_s(t)$ appearing in eq 1 comes from the solution of the differential equation, eq 2, by considering eq 8:

$$v_s(t) = \frac{V_p}{1 + (\Omega\tau_s)^2}\left[\sin(\Omega t + \phi) - \Omega\tau_s\cos(\Omega t + \phi) + \Omega\tau_s e^{-t/\tau_s}\right] \qquad (10)$$

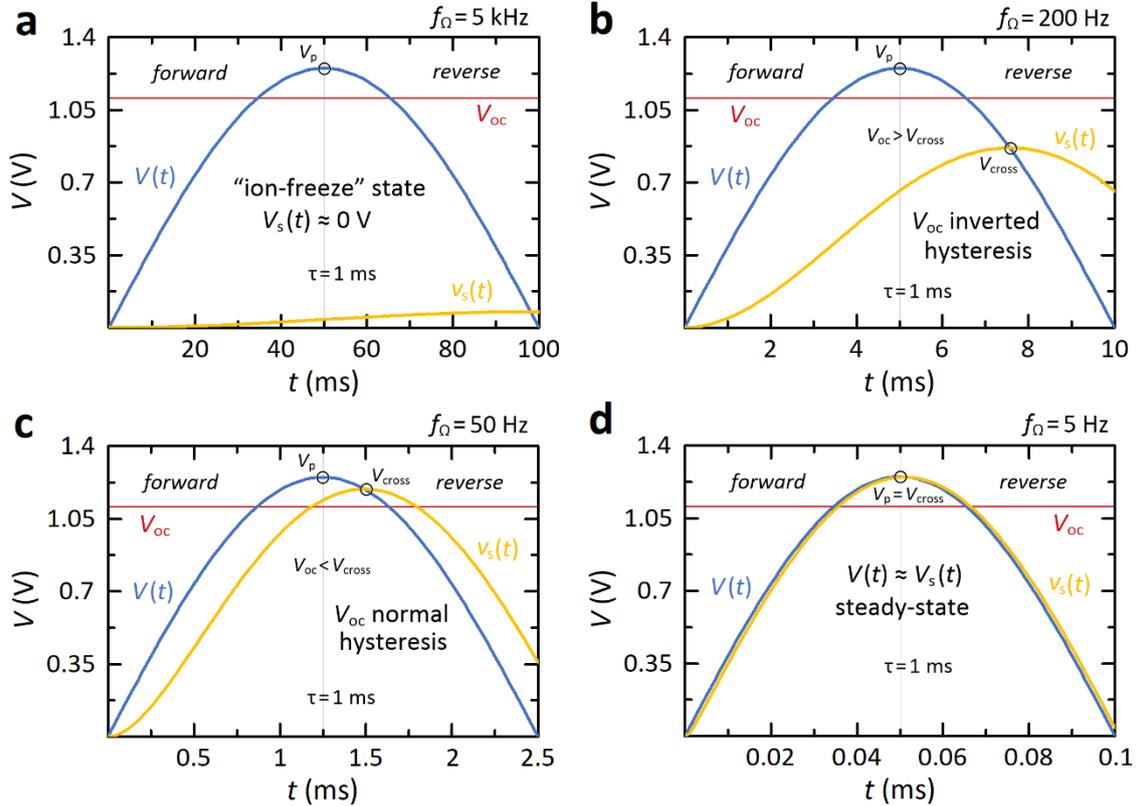

**Figure 1.** Waveforms of the voltages $V(t)$ and $v_s(t)$ corresponding to the perovskite device model in solar cells under sinusoidal sweeps of frequencies $f_\Omega$ (a) 5 kHz, (b) 200 Hz, (c) 50 Hz, and (d) 5 Hz. At high frequencies (a), $v_s(t)$ cannot respond and allows to determine "ion-freeze" efficiency. As the frequency decreases, different types of hysteresis effects appear in the current-voltage curve depending on the crossing of $V(t)$ and $v_s(t)$: if $V_{cross} < V_{oc}$, the hysteresis is predominantly inverted (b) but, if, in contrast, $V_{cross} > V_{oc}$, the hysteresis is normal (c). Finally, both potentials become identical at sufficiently low frequencies as an indicator of hysteresis elimination under steady-state conditions (d). The red colored line indicates the open-circuit voltage value. $V_p = 1.25$ V, $V_0 = 0$ V, $\phi = 0$, and $\tau_s = 1$ ms.

Now, we focus our attention on the evolution of the first two terms of $v_s(t)$ (the last component is a transient exponential-dependent term (depending on the initial condition, $v_s(0) = 0$ V) which dominates the response when $\tau_s > \tau_d$. If the frequency is sufficiently fast ($1 \ll \Omega\tau_s$), $v_s(t)$ tends to zero due to the dominance of the kinetic delay[17] throughout the half-period of the sinusoid. In contrast, hysteresis mechanisms emerge when $1 \sim \Omega\tau_s$ because we obtain an expression for surface potential with additional effects to $V(t)$ that complicates the analysis of the current response (eq 10). Finally, the equilibrium response is not affected and there is no current-voltage hysteresis due to the surface voltage when the scan speed is very slow ($1 \gg \Omega\tau_s$). Next, we summarize the limiting behavior of the surface potential:

$$v_\mathrm{s}(t) = \begin{cases} \dfrac{V_\mathrm{p}\left[\sin(\Omega t + \phi) + \Omega\tau_\mathrm{s}\left(e^{-t/\tau_\mathrm{s}} - \cos(\Omega t + \phi)\right)\right]}{(\Omega\tau_\mathrm{s})^2} \sim 0, & \Omega \gg 1/\tau_\mathrm{s} \\ V_\mathrm{p}\sin(\Omega t + \phi) \sim V(t + \phi), & \Omega \ll 1/\tau_\mathrm{s} \end{cases} \quad (11)$$

By taking the derivative in eq 10, we get

$$\frac{dv_\mathrm{s}(t)}{dt} = \frac{V_\mathrm{p}\Omega}{1 + (\Omega\tau_\mathrm{s})^2}\left[\cos(\Omega t + \phi) + \Omega\tau_\mathrm{s}\sin(\Omega t + \phi) - e^{-t/\tau_\mathrm{s}}\right] \quad (12)$$

to directly include in the calculation of the external current $j$ from eq 1 with a sweep rate corresponding to an angular frequency of $\Omega$ rad/s.

From a direct comparison of eqs 8 and 10, it is evident that the surface potential $v_\mathrm{s}(t)$ exhibits additional terms that introduce a time delay relative to $V(t)$. Depending on the value of the delay of the internal voltage in relation to $f_\Omega$, hysteresis effects will be more or less visible, presenting different features. We apply a voltage schedule with varying sweep rates cycled with frequency and an initial voltage $V_0 = 0$ V (and $\phi = 0$). At ultra-fast voltage scan speeds as indicated in Figure 1a, the surface potential $v_\mathrm{s}(t)$ cannot respond and remains stuck at a negligible value, $v_\mathrm{s}(t) \sim 0$ V. Under this measurement conditions, the ions are immobilized and thus, there is no hysteresis.[28,29] As the frequency decreases, the kinetic delay $\tau_\mathrm{s}$ introduces three regions in the comparison of $V(t)$ and $v_\mathrm{s}(t)$ dominated by capacitive and inductive currents. In the initial and final parts, the time evolution of $V(t)$ and $v_\mathrm{s}(t)$ shows the same trend (increasing and decreasing, respectively), indicative of capacitive hysteresis.[11,30,31] This is the original or normal hysteresis mechanism of halide perovskites, in which the rotation occurs in clockwise sense.

At intermediate time scales, one can however observe a different kind of behavior; when $V(t)$ reaches its maximum value and begins to sinusoidally decays, $v_\mathrm{s}(t)$ continues increasing. This type of behavior leads to counterclockwise loops in current-voltage curves, classified in literature as inverted or inductive hysteresis.[32,33] As previously commented on, this effect emerges because the external and internal voltage are out of phase by exactly $\tau_\mathrm{s}$ seconds. In particular, $V(t)$ leads $v_\mathrm{s}(t)$ by $\Omega\tau_\mathrm{s}$ rad/s, or, equivalently, the surface potential lags behind the external voltage by $\Omega\tau_\mathrm{s}$ rad/s. Consequently, there is a crossing of the curves, suggesting that, exactly at this applied voltage sector ($V_\mathrm{cross}$), a change of current-voltage hysteresis from normal to inverted occurs.[25,34] The dependence of $V_\mathrm{cross}$ with the open-circuit voltage $V_\mathrm{oc}$ (shown as a constant red line in all the panels of Figure 1) plays an important role in the transition of the hysteresis from capacitive to inductive. In Figure 1b, there is a crossing of $V(t)$ and $v_\mathrm{s}(t)$ in a voltage value lower than $V_\mathrm{oc}$, corresponding to the inverted hysteresis. On the other hand, in Figure 1c, we show parameters with $V_\mathrm{cross} > V_\mathrm{oc}$, that produce normal hysteresis. At sufficiently low $f_\Omega$, the phase shift diminishes leading to the disappearance of hysteresis mechanisms under steady-state conditions (Figure 1d). The schematics shown in Figure 1 are a generalization of the hysteretic landscape of perovskite solar cells, in which a mixed hysteresis scenario is assumed and, depending also on the value of the kinetic time constants and the transition voltage value between $V(t)$ and $v_\mathrm{s}(t)$, one or both types will be visualized in the current-voltage curve. In fact, the difference in the values of both potentials here quantifies somewhat the area or index of the hysteresis loop, from which one can extract very relevant physical information of the device operation.[28,35]

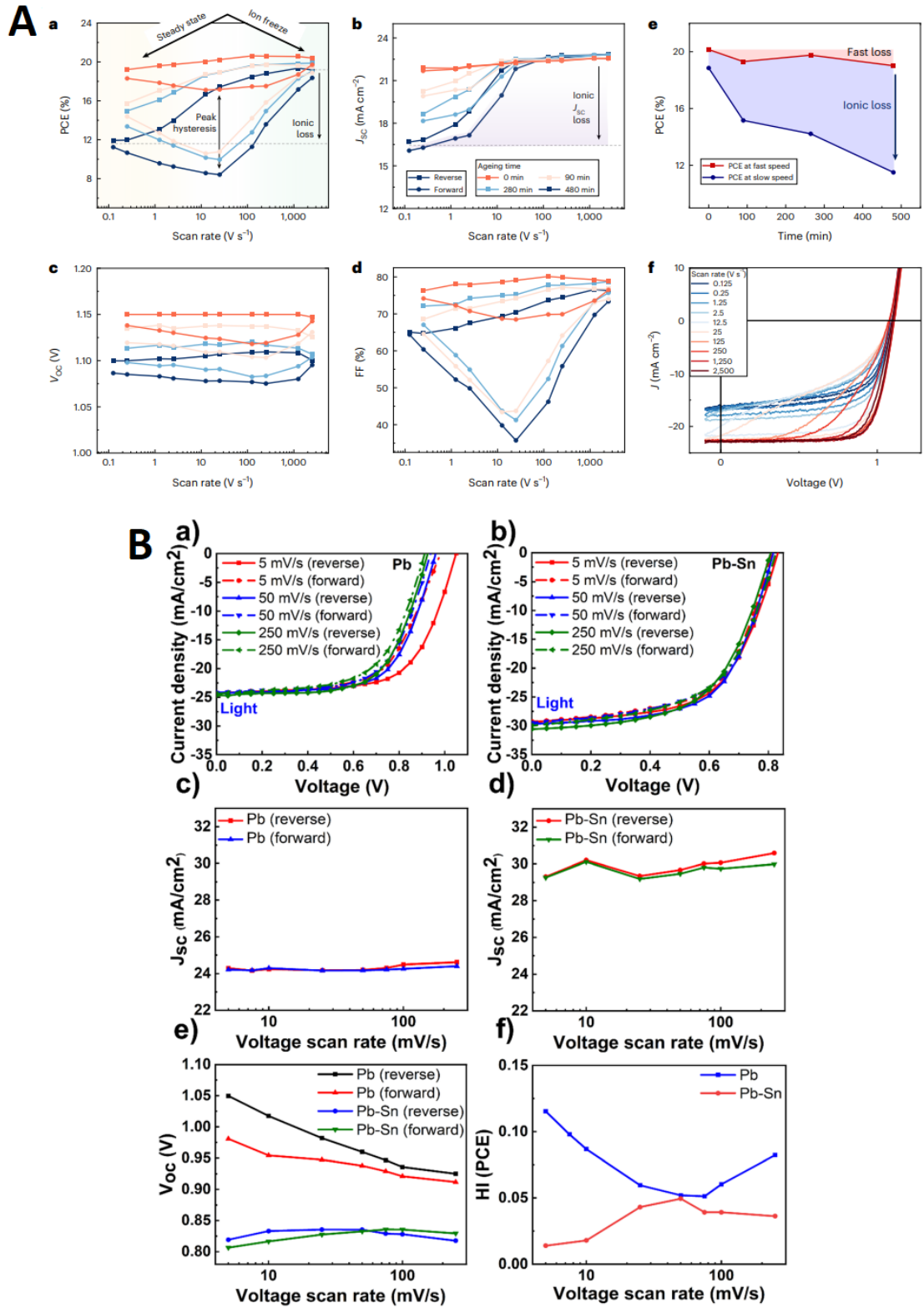

**Figure 2.** (A) η (a), $j_{sc}$ (b), $V_{oc}$ (c), and FF (d) obtained from current-voltage characteristics measured at different scan rates in reverse (squares) and forward (circles) direction for perovskite solar cells based on the nominal formulation $Cs_{0.05}(FA_{0.83}MA_{0.17})_{0.95}Pb(I_{0.83}Br_{0.17})_3$ after different ageing times. Absolute η and representative current-voltage curves after different ageing times. Reproduced from reference 29. Copyright 2024 Springer Nature. (B) Current-voltage measurements of (a) Pb and (b) Pb-Sn perovskite solar cells at three scan rates: 5 mV/s, 50 mV/s, and 250 mV/s. (c) $j_{sc}$ of Pb and (d) Pb-Sn perovskite solar cells as a function of scan rate in the reverse and forward scans. (e) $V_{oc}$ of Pb and Pb-Sn perovskite devices

7as a function of scan rate in the reverse and forward scans. (f) HI in η of Pb and Pb-Sn perovskite solar cells as a function of scan rate. Reproduced from reference 37. Copyright 2024 Royal Society of Chemistry.

Analogously to the analysis of the delayed surface voltage, the differential equation (eq 3) can be solved for the initial condition $j_d(0) = 0$ A to provide the expression of the ion-modulated current $j_d(t)$[25] in time domain:

$$j_d(t) = \frac{J_{elect0}}{1 + \tau_d \frac{q}{n_d k_B T} \frac{dV(t)}{dt}} \left[ e^{\frac{qV}{n_d k_B T}} - e^{-t/\tau_d} \right] \quad (13)$$

where

$$\frac{dV(t)}{dt} = V_p \Omega \cos(\Omega t + \phi) \quad (14)$$

also useful for the displacement current that charges $C_g$ eq 1. Note that $j_d$ governs the overall response when $\tau_d > \tau_s$. The approach given by eq 13 leads to the same conclusions that those for $v_s(t)$ in eq 11:

$$j_d(t) = \begin{cases} J_{elect0} e^{\frac{qV}{n_d k_B T}} \sim J_{elect}(V), & \Omega \ll 1/\tau_s \\ \frac{J_{elect0} n_d k_B T}{\tau_d q V_p \Omega \cos(\Omega t + \phi)} \left[ e^{\frac{qV}{n_d k_B T}} - e^{-t/\tau_d} \right], & \Omega \gg 1/\tau_s \end{cases} \quad (15)$$

depending on the voltage scan speed. Before proceeding further, it is important to point out that, in any case, electrical measurements indicate that the time constants $\tau_s$ and $\tau_d$ are similar ($\tau_s \sim \tau_d$) but not exactly the same as dictated the original surface polarization model[21,36] and, depending on which is larger and thus dominant in the response, different features are observed in the experimental data.[19]

At this point, it remains latent the following question: What is the variation of photovoltaic parameters as a function of the frequency? This type of representations are widely reported in the literature[28,29,35,37] and we include, in this regard, a selection of representative examples, obtained from experimental measurements, in Figure 2, where the power conversion efficiencies (η), short-circuit current densities ($j_{sc}$), fill factors (FF) and open-circuit voltages ($V_{oc}$) in forward and reverse scans are plotted in a wide range of scan rates. In effect, it has been recognized that the analysis of photovoltaic parameters with respect to the frequency/scan rate provides important information about the dynamic behavior of the perovskite devices. Specifically, it is observed in Figure 2A that hysteresis first increases and then decreases at ultra-fast scan speeds when the movement of ions is "frozen" and, consequently, a concomitant increase of η is obtained as it was remarked in the reference 29. Here, the hysteresis is obtained predominantly close to $V_{oc}$, as a result of degradation of the device performance.[37] The initial cell also shows invariable $j_{sc}$ at forward and reverse directions, but when the device is degraded significant deviations occur in $j_{ph}$, due to degradation causing charge collection issues and electrical field dependence. Similarly, Figure 2B shows results for high performance perovskite solar cells. An important result observed in this work is that $V_{oc}$ hysteresis in Pb cell tends to decrease at high scan rates. Another significant behavior is observed in the Pb-Sn device. Here, $V_{oc}$ is changing sign when the scan rate increases. It means that the hysteresis is first normal and then inverted; i.e., there is a transition of hysteresis as the scan rate increases, which causes a change of the tendency of the variations of η. Next, we aim to explain the typical evolution of hysteresis observed in the current-voltage curves shown in Figure 2, from sinusoidal voltage sweeps and corresponding to the change of the impedance from capacitive to inductive response.

For a further interpretation of the complex dynamical properties of current-voltage responses in perovskite solar cells, we first develop eqs 1−3 into the corresponding small perturbation equations in the time-domain,[18,19] yielding

$$\Delta j(t) = C_g \frac{d}{dt}\Delta V + \Delta V g_{rec} + j_d + C_s \frac{d}{dt}\Delta v_s(t) \tag{16}$$

$$\tau_s \frac{d}{dt}\Delta v_s(t) = \Delta V - \Delta v_s(t) \tag{17}$$

$$\tau_d \frac{d}{dt}\Delta j_d(t) = \Delta V g_{elect} - j_d \tag{18}$$

from the first-order Taylor expansion around the steady-state values. The analytical solution of the state function given in eqs 17 and 18, by considering $\Delta V(t) = \Delta V$ (with $\Delta V > 0$; i.e., forward voltage sweep), may be expressed as

$$\Delta v_s(t) = \Delta V \left(1 - e^{-t/\tau_s}\right) \tag{19}$$
$$\Delta j_d(t) = \Delta V g_{elect}\left(1 - e^{-t/\tau_d}\right) \tag{20}$$

which somewhat constitutes the last two channels in eq 16 that represent the slow terms of eq 16. From eqs 19 and 20, we obtain a reasonable approximation of the complete transient response $\Delta j(t)$ due to a small perturbation of voltage $\Delta V$ in perovskite solar cells, defined as:

$$\Delta j(t) = \Delta j_{rec}(V) + \Delta V g_{ion} e^{-t/\tau_s} - \Delta V g_{elect} e^{-t/\tau_d} \tag{21}$$

By considering that both time constants are approximately the same as commented previously ($\tau_s \sim \tau_d$) called $\tau_{kin}$ in previous publications, the total current of eq 16 can be therefore simplified to[18]

$$\Delta j(t) = \Delta j_{rec}(V) + \Delta V (g_{ion} - g_{elect}) e^{-t/\tau_{kin}} \tag{22}$$

where the two conductances $g_{ion}$ and $g_{elect}$ are defined as

$$g_{ion} = \frac{1}{\tau_s}\frac{dQ_s}{dv_s} = \frac{C_s}{\tau_s} \tag{23}$$

$$g_{elect} = \frac{dJ_{elect}}{dv_s} = \frac{q}{n_d K_B T} J_{elect}(V) \tag{24}$$

differing in the pre-factor and slightly in the voltage dependence because $n_{rec} \sim n_s \sim n_d$. Both terms represent the essential mechanism for the electronic recombination controlled by ionic motion at long time scales in metal halide perovskites, associated with negative capacitance effects.[38] Importantly, it is necessary to point out that there is an additional conductance in the model,

$$g_{rec} = \frac{dJ_{rec}}{dV} = \frac{q}{n_{rec} K_B T} J_{rec}(V) \tag{25}$$

that constitutes the fast step of electronic current recombination. Thus,

$$\Delta j_{rec}(\Delta V) = \Delta V (g_{rec} + g_{elect}) \tag{26}$$

that is, the current value when the dc situation is achieved. Note that "Δ" represents small displacements, so that all the variables remain close to the equilibrium.

The analysis of the ionic-electronic response in the time domain is of special interest since eq 22 shows that, when a potential step is applied, the perovskite device returns to equilibrium increasing or decreasing exponentially depending on the value of the conductances given by eqs 23 and 24.[19] In effect, the sign of the memory-based currents in eq 22 plays a pivotal role in the transition of the current-voltage hysteresis from capacitive to inductive, depending on the dominant conductance. If $g_{ion} > g_{elect}$, then the forward current is positive, corresponding to capacitive hysteresis. In contrast, the additional current is negative when $g_{ion} < g_{elect}$ in the forward voltage sweep, decreasing the operational current level and thus forming an inductive hysteresis loop.[39] The reasoning behind this mathematical approach can be extrapolated to backward voltage scanning, when $\Delta V < 0$.

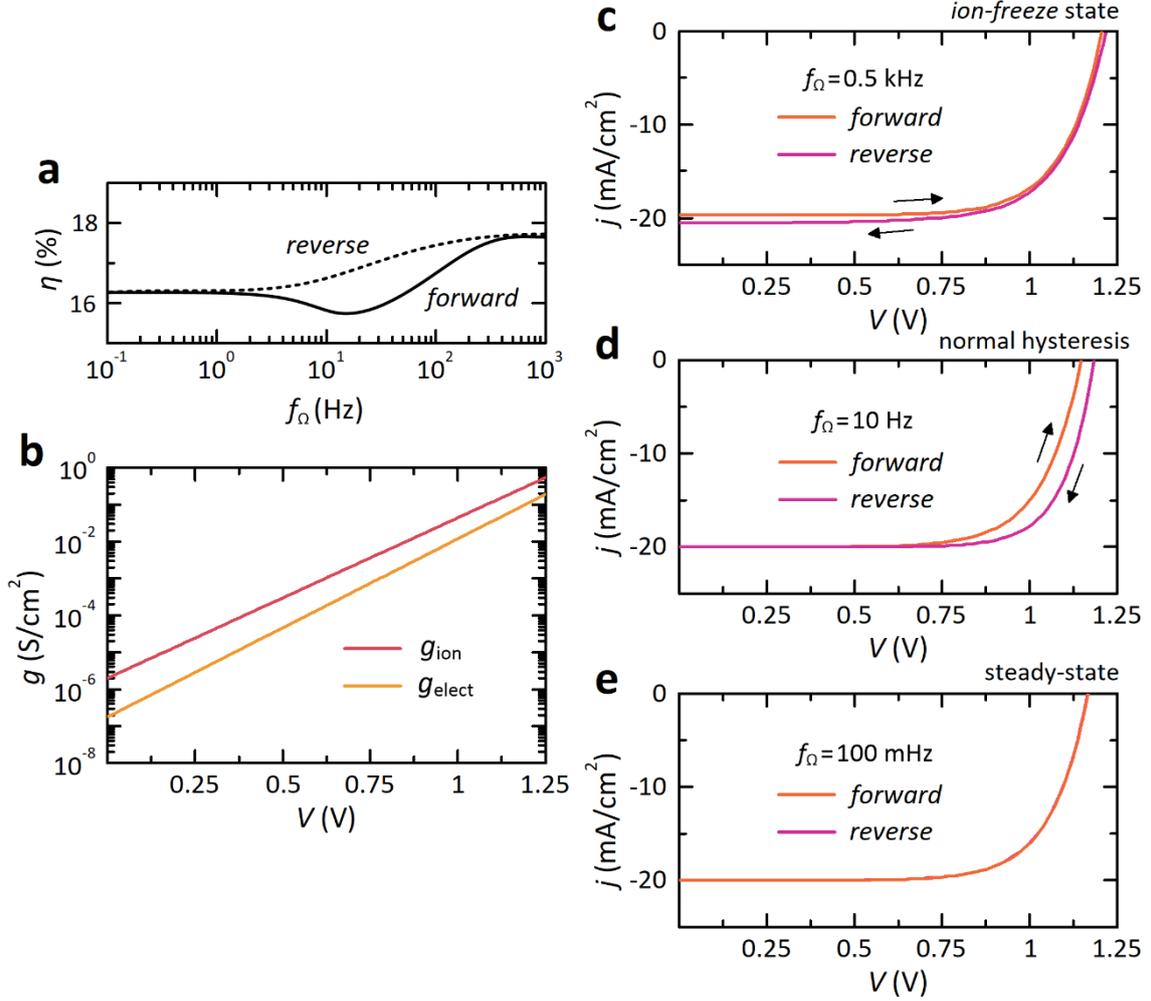

**Figure 3.** (a) Approximated evolution of the power conversion efficiency η as a function of the frequency $f_\Omega$ for the simulated current-voltage characteristics shown in Figures 3c–e, respectively. (b) Simulated conductance-voltage profiles and the corresponding current-voltage curves in forward and reverse directions for sinusoidal scans of different frequency: (c) 0.5 kHz, (d) 10 Hz, and (e) 100 mHz. Parameters: $V_p = 1.25$ V, $k_B T/q = 0.026$ V, $C_g = 0.1$ μF/cm², $j_{rec0} = 0.33$ μA/cm², $n_{rec} = 4.23$, $C_{s0} = 2$ nF/cm², $n_s = 3.84$, $j_{elect0} = 16$ nA/cm², $n_d = 3.46$, $\tau_s = \tau_d = 1$ ms, and $j_{ph} = 20$ mA/cm². In the simulations, we assume negligible series resistance effects.

Reproducing the results of the literature through our mathematical model, Figure 3a shows the approximated evolution of η as a function of the cycled frequency $f_\Omega$ in which one can easily deduce that capacitive effects dominate the response throughout the current-voltage curve (as in Figure 2A) because $\eta_{reverse} > \eta_{forward}$ always. We initially represent the associated conductance responses as a function of the applied voltage without, in effect, a crossing voltage that leads to a transition from capacitive to inductive effects (Figure 3b). Furthermore, we assume $\tau_s = \tau_d$. A summary of the pertinent features of the current-voltage curves simulated from our theory, derived by one of the mathematical approaches belonging to the family of fast-slow perovskite models of literature,[17] is shown in the catalogue of simulated responses shown in Figure 3c–e. It is observed that Figures 3c and 3e, obtained at sufficiently fast and slow sinusoidal sweeps, respectively, do not show current-voltage hysteresis, corresponding such curves to the "ion-freeze" and steady-state efficiencies that coincide under the forward and backward scan direction,[27,29] as indicated in Figure 3a. On the other hand, capacitive (normal) behavior of current-voltage hysteresis is shown at intermediate frequencies in Figure 3d, also known as peak of hysteresis, where it emerges the maximum difference between η determined from the forward versus the reverse scan.

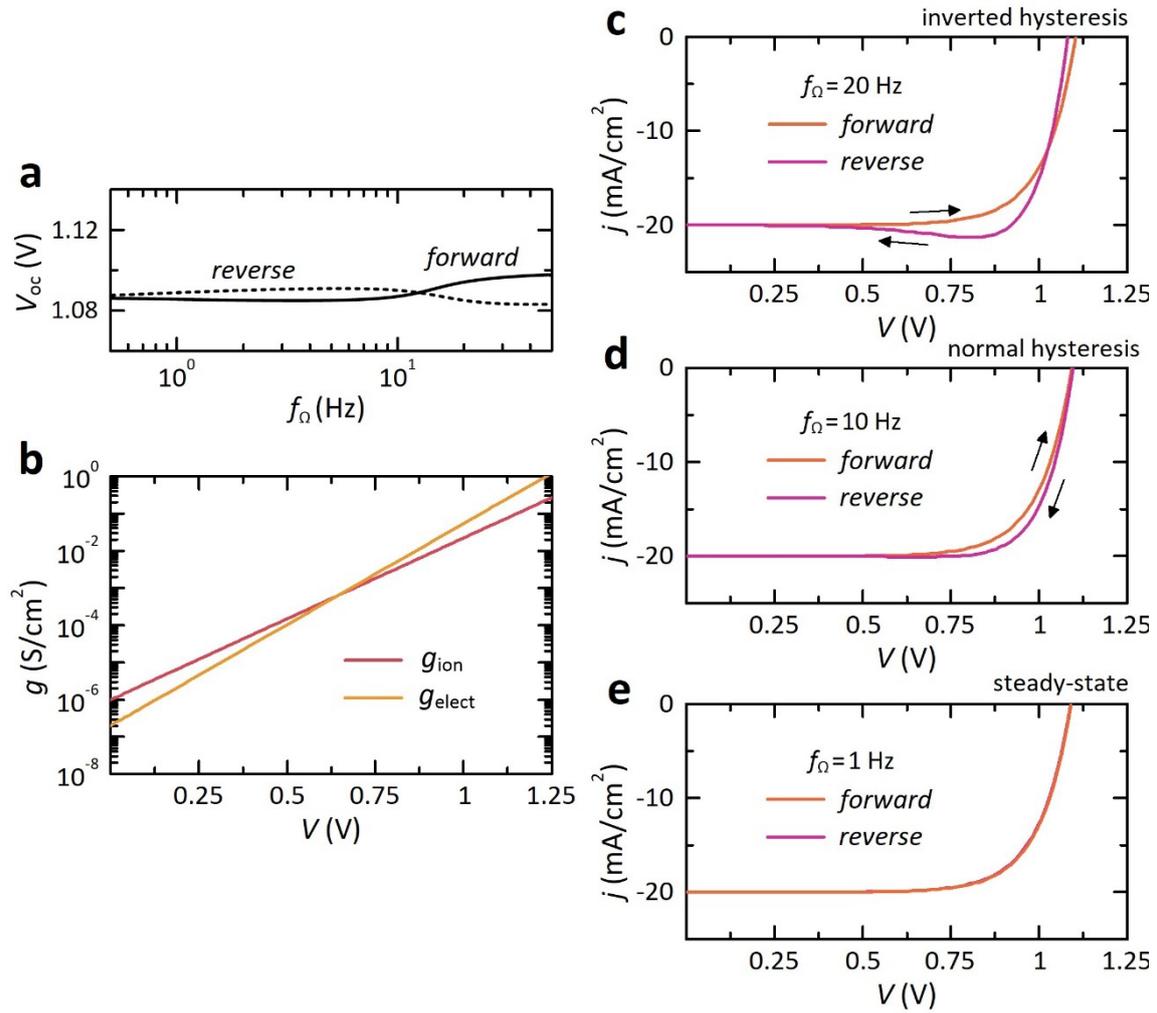

**Figure 4.** (a) Approximated evolution of the open-circuit voltage $V_{oc}$ as a function of intermediate/slow frequencies $f_\Omega$ for the simulated current-voltage curves shown in Figures 4c–e, respectively. (b) Simulated conductance-voltage profiles varying the voltage regime where the ionic-electronic components dominate, and the corresponding current-voltage curves in forward and reverse directions for sinusoidal scans of different frequency: (c) 20 Hz, (d) 10 Hz, and (e) 1 Hz. Crossing voltage in (b), together with the value of $f_\Omega$ (explanation of Figure 1), plays a key role in the nature of hysteresis; inductive or inverted in (c) and capacitive or regular in (d) around $V_{oc}$. Parameters: $V_p = 1.25$ V, $k_B T/q = 0.026$ V, $C_g = 0.1$ μF/cm², $j_{rec0} = 0.33$ μA/cm², $n_{rec} = 4.23$, $C_{s0} = 2$ nF/cm², $n_s = 3.84$, $\tau_s = 2$ ms, $j_{elect0} = 16$ nA/cm², $n_d = 3.07$, $\tau_d = 0.5$ ms, and $j_{ph} = 20$ mA/cm². In the simulations, we assume negligible series resistance effecta.

Next, we explain another relevant feature of the hysteresis consisting of the transition from normal to inverted, as observed in Figure 2B. From numerical simulations, we reproduce such behavior in which a transformation in the dominant mechanism of $V_{oc}$ hysteresis from capacitive ($V_{oc, reverse} > V_{oc, forward}$) to inductive ($V_{oc, reverse} < V_{oc, forward}$) emerges as frequency increases (refer to Figure 4a). It is important to clarify that inductive effects are typically found in a mixed hysteresis scenario, leading to sign changes in, for example, the hysteresis index (HI).[40] All these properties are outlined from the physical landscape of voltage-dependent ionic/electronic conductivity in perovskite solar cells. Figure 4b shows the evolution of the two slow conductances, where it now appears a crossing point and thus, a corresponding change of the type of hysteresis because we also assume $\tau_s < \tau_d$. In effect, the general partition of types of hysteresis shown in Figures 4c–e is well-supported by the previous conductance data. We can see that the crossing point depends on the intrinsic electrical parameters related to $g_{ion}$ and $g_{elect}$. By eqs 23 and 24, we can obtain the voltage value associated with the crossing $V_{cross}$ due to the change of dominant conductance:

$$V_{cross} = n_d \left( \frac{V_s}{n_s} + \frac{k_B T}{q} \ln \left[ \frac{C_{s0} n_d k_B T}{\tau_s q J_{elect0}} \right] \right) \tag{27}$$

where, if $V_{cross} > V_{oc}$ or $V_{cross} \sim V_{oc}$, the type of hysteresis is capacitive –Figure 4d– and if, in contrast, $V_{cross} < V_{oc}$, the prevalent mechanism is of chemical inductive nature in the current-voltage curve around $V_{oc}$ (inverted hysteresis, refer to Figure 4c). As it is well-known in the literature, there is not logically hysteresis in the current-voltage curves simulated at slow scan speeds (equilibrium conditions, refer to Figure 4e). Interpretation and evolution of the key parameters in eq 27 governing the general shapes of the hysteresis has been amply discussed for the perovskite's community from the perspective of equivalent circuits via Impedance Spectroscopy. Regarding the ideality factor, it is typical to find resistances with similar voltage or illumination dependence, suggesting a common kinetic mechanism of origin.[23,41] Nevertheless, this situation is far from general as physical conditions, device architecture and sample stability influence the experimental results.[42,43] On the other hand, it is important to note that we use a simple approach for numerical simulations in which the memory variable $v_s$ exhibits a constant relaxation time $\tau_s$, as shown the results of literature.

To carefully assess the significant impact of the conductances, as well as the frequency, in the hysteresis phenomenology, we next analyze the natural framework of Impedance Spectroscopy.[44] From the introduction of phasors to eqs 16, 17, and 18,

$$\hat{j} = j\omega C_g \hat{V} + g_{rec}\hat{V} + \hat{j_d} + j\omega C_s \hat{v_s} \tag{28}$$

$$\hat{v_s} = \frac{\hat{V}}{1 + j\omega\tau_s} \tag{29}$$

$$\hat{j_d} = \frac{\hat{V} g_{elect}}{1 + j\omega\tau_d} \tag{30}$$

we find the expression of the impedance[18,36]

$$Z(\omega) = \frac{\hat{V}}{\hat{j}} = \left( j\omega C_g + g_{rec} + \frac{1}{\frac{1}{j\omega C_s} + \frac{1}{g_{ion}}} + \frac{1}{\frac{1}{g_{elect}} + j\omega L_d} \right)^{-1} \tag{31}$$

where the diacritic "^" denotes small ac perturbations of angular frequency $\omega$, added to the steady-state level. Note that the effect of geometrical capacitance $C_g$, negligible for hysteresis effects, here represent the dielectric bulk processes that are visible at high frequencies. The last two terms of the right-hand side of eq 31 represents two different conduction pathways controlled by the similar relaxation times, $\tau_s$ and $\tau_d$,[45,46] as corroborated by the experimental results of the literature.[18,25,38,47] The term $\left(1/j\omega C_s + 1/g_{ion}\right)^{-1}$ denotes an RC line with $R_{ion} = 1/g_{ion}$ and $C_s$. On the other hand, the electrical approach of the mathematical component $\left(1/g_{elect} + j\omega L_d\right)^{-1}$ is a series RL branch, where the resistance is $R_{elect} = 1/g_{elect}$ and the inductance is expressed as

$$L_d = \frac{\tau_d}{g_{elect}} \tag{32}$$

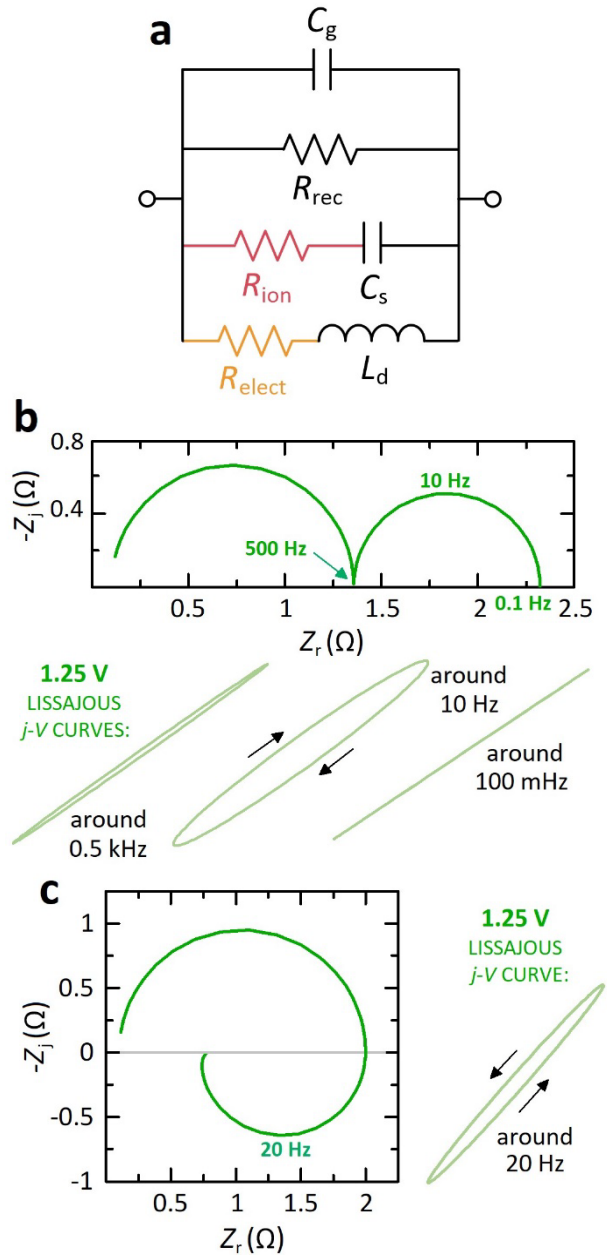

**Figure 5.** (a) Equivalent circuit found from the linearization of the analytical equations of the surface polarization model in photovoltaic perovskites. From the value of the parameters used in the numerical

simulations of Figure 3, (b) impedance is capacitive at low frequencies, the current-voltage Lissajous curve is clockwise and there are free-hysteresis curves at much higher and lower frequencies. In contrast, (c) the impedance with the value of the parameters used in Figure 4 exhibits inductive features at sufficiently low frequencies and the hysteresis describes a counterclockwise loop in the corresponding Lissajous curve and the current-voltage curves, showing the general connection between the impedance components and the hysteresis behavior.

Eq. (29) is represented as an equivalent circuit in Figure 5a. This electrical model, first derived from the surface polarization model[21], produces a chemical inductor structure[48] that evidences crucial processes in perovskite semiconductors, in terms of delayed electronic effects via ionic motion, which manifest themselves through apparent negative capacitances,[19,36] negative transient spikes,[39,49] and/or inverted hysteresis.[33,38] All this phenomenology is ultimately deleterious for the photovoltaic performance,[50] but instead represents a central part of the functionality in memristors for neuromorphic computing.[51]

From impedance analysis, we next illustrate the relationship between the hysteresis mechanisms and the sweep velocity of current-voltage curves, according to device electrical properties (capacitive or inductive) at the frequency under study. The capacitive spectrum shown in Figure 5b, consisting of two arcs in the low-voltage region, corresponds to the regular or normal hysteresis of Figure 3d. Inductive effects, in contrast, emerge at sufficiently low frequencies with the same parameter values of those of current-voltage curves obtained in Figure 4c−e. It causes a negative transient behavior in the currents, as precited in eq 22, and finally resulting in an inverted hysteresis around $V_{oc}$ – refer to Figure 4c–. The insets of Figures 5b and 5c show the respective Lissajous curves corresponding to the voltage-current representation in impedance found at the specific frequencies[28,52] at which the current-voltage responses of Figures 3 and 4 were obtained. The current describes different clockwise and counterclockwise loops for low-frequency capacitive and inductive behavior in impedance, predicting the type and level of hysteresis in the current-voltage curves.[53] On the other hand, the Lissajous curves, which are basically straight lines in Figure 5b, correspond to the current-voltage relationships at certain frequencies where the electrical behavior of the perovskite is purely resistive. Approximately at such frequencies, that describe the beginning and end of the second capacitive arc shown in Figure 5b at low frequencies, there is no hysteresis, either because it is in the state of ion-freeze or equilibrium corresponding to Figures 3c and 3e, respectively. Note that a capacitive current-voltage Lissajous loop is not found at lower frequencies in Figure 5c because there is not practically hysteresis around $V_{oc}$ in Figure 4d. Although it exists a clear correspondence between impedance and current-voltage experiments operated here both under sinusoidal operation, the natural framework of these measurements is fully different; that is, the connection between the small and large perturbation techniques is not direct.

**Conclusions**

In conclusion, although the perovskite device efficiencies have been widely reported via a range of widely used characterization measurement protocols with potentiostatic stimulus exhibiting different forms,[6,11,54,55] the ambiguity that hysteresis causes in such photovoltaic devices dictates that new horizons should be devised in analyzing the performance of metal halide perovskites. Instead of the traditional way of analyzing current-voltage curves by linear sweep voltammetry, we here disentangle the hysteresis phenomenology of simulated nonlinear responses in the frequency domain using large sinusoidal excitations to directly correlate with impedance responses. Specifically, we study these anomalous mechanisms depending on scan voltage frequency and the significant impact it has on the alteration of the curve shape and the evolution of photovoltaic parameters. We established a general connection between the dominant element of impedance, the frequency in which it emerges, and the type of hysteresis loop is obtained under different voltage sweep velocities even though these techniques are not unequivocally linked because they operate under different linearity conditions. The advanced analytical equations, derived by our physics-based perovskite device model, introduce the possibility of quantitatively explore the evolution of device efficiencies in a reliable way at very low data management cost, as a great promise for the characterization of perovskite solar cells in large-scale experimental protocols.

## Associated content

*Supporting information.* Mathematical approach for current-voltage curves in perovskites obtained with a non-zero bias point and changing the directionality of the sweep.

*Data availability statement.* The data presented here can be accessed at https://doi.org/10.5281/zenodo.12611103 (Zenodo) under the license CC-BY-4.0 (Creative Commons Attribution-ShareAlike 4.0 International).

## Conflict of interests

There are no conflicts to declare.

## Acknowledgements

This work has received funding from the Universidad Rey Juan Carlos, project number M2993. This work was funded by the European Research Council (ERC) via Horizon Europe Advanced Grant, grant agreement nº 101097688 ("PeroSpiker").

## References

1   M. A. Green, A. Ho-Baillie and H. J. Snaith, *Nature Photon.*, 2014, **8**, 506–514.
2   J. P. Correa-Baena, M. Saliba, T. Buonassisi, M. Grätzel, A. Abate, W. Tress and A. Hagfeldt, *Science* 2017, **358** (6364), 739–744.
3   J. Y. Kim, J.-W. Lee, H. S. Jung, H. Shin and N.-G. Park, *Chem. Rev.*, 2020, **120**, 7867–7918.
4   H. J. Snaith, A. Abate, J. M. Ball, G. E. Eperon, T. Leijtens, N. K. Noel, S. D. Stranks, J. T.-W. Wang, K. Wojciechowski and W. Zhang, *J. Phys. Chem. Lett.*, 2014, **5**, 1511–1515.
5   E. L. Unger, E. T. Hoke, C. D. Bailie, W. H. Nguyen, A. R. Bowring, T. Heumüller, M. G. Christoforo and M. D. McGehee, *Energy Environ. Sci.*, 2014, **7**, 3690–3698.
6   W. Tress, N. Marinova, T. Moehl, S. M. Zakeeruddin, M. K. Nazeeruddin and M. Grätzel, *Energy Environ. Sci.*, 2015, **8**, 995–1004.
7   S. van Reenen, M. Kemerink and H. J. Snaith, *J. Phys. Chem. Lett.*, 2015, **6**, 3808–3814.
8   S. Meloni, T. Moehl, W. Tress, M. Franckevičius, M. Saliba, Y. H. Lee, P. Gao, M. K. Nazeeruddin, S. M. Zakeeruddin, U. Rothlisberger and M. Graetzel, *Nat. Commun.*, 2016, **7**, 1–9.
9   P. Calado, A. M. Telford, D. Bryant, X. Li, J. Nelson, B. C. O'Regan and P. R. F. Barnes, *Nat. Commun.*, 2016, **7**, 13831.
10  R. S. Sanchez, V. Gonzalez-Pedro, J.-W. Lee, N.-G. Park, Y. S. Kang, I. Mora-Sero and J. Bisquert, *J. Phys. Chem. Lett.*, 2014, **5**, 2357–2363.
11  B. Chen, M. Yang, X. Zheng, C. Wu, W. Li, Y. Yan, J. Bisquert, G. Garcia-Belmonte, K. Zhu and S. Priya, *J. Phys. Chem. Lett.*, 2015, **6**, 4693–4700.

12   J. A. Christians, J. S. Manser and P. V. Kamat, *J. Phys. Chem. Lett.*, 2015, **6**, 852–857.
13   C. Liu, P. J. Tiw, T. Zhang, Y. Wang, L. Cai, R. Yuan, Z. Pan, W. Yue, Y. Tao and Y. Yang, *Nat. Commun.*, 2024, **15**, 1523.
14   X. Zhou, Y. Zong, Y. Wang, M. Sun, D. Shi, W. Wang, G. Du and Y. Xie, *Natl. Sci. Rev.* 2024, **11** (4), nwad216.
15   L. Chua, *Appl. Phys. A*, 2011, **102**, 765–783.
16   W. Yi, K. K. Tsang, S. K. Lam, X. Bai, J. A. Crowell and E. A. Flores, *Nat. Commun.*, 2018, **9**, 4661.
17   J. Bisquert, *Adv. Energy Mater.* 2024, doi.org/10.1002/aenm.202400442
18   E. H. Balaguera and J. Bisquert, *ACS Energy Lett.* 2024, **9** (2), 478–486.
19   E. Hernández-Balaguera and J. Bisquert, *Adv. Funct. Mater.* 2023, **34** (6), 2308678.
20   G. A. Nemnes, C. Besleaga, A. G. Tomulescu, I. Pintilie, L. Pintilie, K. Torfason and A. Manolescu, *Sol. Energy Mater. Sol. Cells*, 2017, **159**, 197–203.
21   S. Ravishankar, O. Almora, C. Echeverría-Arrondo, E. Ghahremanirad, C. Aranda, A. Guerrero, F. Fabregat-Santiago, A. Zaban, G. Garcia-Belmonte and J. Bisquert, *J. Phys. Chem. Lett.*, 2017, **8**, 915–921.
22   R. Gottesman, P. Lopez-Varo, L. Gouda, J. A. Jimenez-Tejada, J. Hu, S. Tirosh, A. Zaban and J. Bisquert, *Chem*, 2016, **1**, 776–789.
23   I. Zarazua, G. Han, P. P. Boix, S. Mhaisalkar, F. Fabregat-Santiago, I. Mora-Seró, J. Bisquert and G. Garcia-Belmonte, *J. Phys. Chem. Lett.*, 2016, **7**, 5105–5113.
24   H.-S. Kim, I.-H. Jang, N. Ahn, M. Choi, A. Guerrero, J. Bisquert and N.-G. Park, *J. Phys. Chem. Lett.* 2015, **6** (22), 4633–4639.
25   C. Gonzales, A. Guerrero and J. Bisquert, *J. Phys. Chem. C*, 2022, **126**, 13560–13578.
26   E. Hernández-Balaguera, L. Muñoz-Díaz, C. Pereyra, M. Lira-Cantú, M. Najafi and Y. Galagan, *Mater. Today Energy*, 2022, **27**, 101031.
27   J. Bisquert, A. Guerrero and C. Gonzales, *ACS Phys. Chem Au*, 2021, **1**, 25–44.
28   V. M. Le Corre, J. Diekmann, F. Peña-Camargo, J. Thiesbrummel, N. Tokmoldin, E. Gutierrez-Partida, K. Pawel Peters, L. Perdigón-Toro, M. H. Futscher, F. Lang, J. Warby, H. J. Snaith, D. Neher and M. Stolterfoht, *Sol. RRL* 2022, **6** (4), 2100772.
29   J. Thiesbrummel, S. Shah, E. Gutierrez-Partida, F. Zu, F. Peña-Camargo, S. Zeiske, J. Diekmann, F. Ye, K. P. Peters, K. O. Brinkmann, P. Caprioglio, A. Dasgupta, S. Seo, F. A. Adeleye, J. Warby, Q. Jeangros, F. Lang, S. Zhang, S. Albrecht, T. Riedl, A. Armin, D. Neher, N. Koch, Y. Wu, V. M. Le Corre, H. Snaith and M. Stolterfoht, *Nat. Energy*, 2024, 1–13.
30   O. Almora, I. Zarazua, E. Mas-Marza, I. Mora-Sero, J. Bisquert and G. Garcia-Belmonte, *J. Phys. Chem. Lett.*, 2015, **6**, 1645–1652.
31   E. Hernández-Balaguera, G. del Pozo, B. Arredondo, B. Romero, C. Pereyra, H. Xie and M. Lira-Cantú, *Solar RRL*, 2021, **5**, 2000707.
32   W. Tress, J. P. C. Baena, M. Saliba, A. Abate and M. Graetzel, *Adv. Energy Mater.*, 2016, **6**, 1600396.
33   A. O. Alvarez, R. Arcas, C. A. Aranda, L. Bethencourt, E. Mas-Marzá, M. Saliba and F. Fabregat-Santiago, *J. Phys. Chem. Lett.*, 2020, **11**, 8417–8423.
34   G. A. Nemnes, C. Besleaga, V. Stancu, D. E. Dogaru, L. N. Leonat, L. Pintilie, K. Torfason, M. Ilkov, A. Manolescu and I. Pintilie, *J. Phys. Chem. C*, 2017, **121**, 11207–11214.
35   J. Wu, Y. Li, Y. Li, W. Xie, J. Shi, D. Li, S. Cheng and Q. Meng, *J. Mater. Chem. A*, 2021, **9**, 6382–6392.

36   E. Ghahremanirad, A. Bou, S. Olyaee and J. Bisquert, *J. Phys. Chem. Lett.*, 2017, **8**, 1402–1406.
37   K. Dey, D. Ghosh, M. Pilot, S. R. Pering, B. Roose, P. Deswal, S. P. Senanayak, P. J. Cameron, M. Saiful Islam, S. D. Stranks, *Energy Environ. Sci.* **2024**, 17, 760–769.
38   F. Ebadi, N. Taghavinia, R. Mohammadpour, A. Hagfeldt and W. Tress, *Nat, Commun,*, 2019, **10**, 1574.
39   E. Hernández-Balaguera and J. Bisquert, *ACS Energy Lett.*, 2022, **7**, 2602–2610.
40   S. N. Habisreutinger, N. K. Noel and H. J. Snaith, *ACS Energy Lett.*, 2018, **3**, 2472–2476.
41   O. Almora, K. T. Cho, S. Aghazada, I. Zimmermann, G. J. Matt, C. J. Brabec, M. K. Nazeeruddin and G. Garcia-Belmonte, *Nano Energy*, 2018, **48**, 63–72.
42   P. Lopez-Varo, J. A. Jiménez-Tejada, M. García-Rosell, S. Ravishankar, G. Garcia-Belmonte, J. Bisquert and O. Almora, *Adv. Energy Mater.*, 2018, **8**, 1702772.
43   E. Hernández-Balaguera, B. Romero, B. Arredondo, G. del Pozo, M. Najafi and Y. Galagan, *Nano Energy*, 2020, **78**, 105398.
44   A. Guerrero, J. Bisquert and G. Garcia-Belmonte, *Chem. Rev.*, 2021, **121**, 14430–14484.
45   J. Bisquert, *J. Phys. Chem. Lett.*, 2023, **14**, 1014–1021.
46   N. Filipoiu, A. T. Preda, D.-V. Anghel, R. Patru, R. E. Brophy, M. Kateb, C. Besleaga, A. G. Tomulescu, I. Pintilie, A. Manolescu and G. A. Nemnes, *Phys. Rev. Appl.*, 2022, **18**, 064087.
47   E. Hernández-Balaguera and D. Martín-Martín, *Fractal Fract.* 2023**, 7** (7), 516.
48   J. Bisquert and A. Guerrero, *J. Am. Chem. Soc.*, 2022, **144** (13), 5996–6009.
49   S. E. J. O'Kane, G. Richardson, A. Pockett, R. G. Niemann, J. M. Cave, N. Sakai, G. E. Eperon, H. J. Snaith, J. M. Foster, P. J. Cameron and A. B. Walker, *J. Mater. Chem. C*, 2017, **5** (2), 452–462.
50   F. Fabregat-Santiago, M. Kulbak, A. Zohar, M. Vallés-Pelarda, G. Hodes, D. Cahen and I. Mora-Seró, *ACS Energy Lett.* 2017, **2** (9), 2007–2013.
51   E. Hernández-Balaguera, L. Munoz-Díaz, A. Bou, B. Romero, B. Ilyassov, A. Guerrero and J. Bisquert, *Neuromorphic Comput. Eng.* 2023, **3** (2), 024005.
52   N. Pellet, F. Giordano, M. Ibrahim Dar, G. Gregori, S. M. Zakeeruddin, J. Maier, and M. Grätzel, *Prog. Photovol.* 2017, **25**, 942-950.
53   J. Bisquert, *PRX Energy*, 2024, **3**, 011001.
54   B. Chen, M. Yang, S. Priya and K. Zhu, *J. Phys. Chem. Lett.*, 2016, **7**, 905–917.
55   N. S. Hill, M. V. Cowley, N. Gluck, M. H. Fsadni, W. Clarke, Y. Hu, M. J. Wolf, N. Healy, M. Freitag, T. J. Penfold, G. Richardson, A. B. Walker, P. J. Cameron and P. Docampo, *Adv. Mater.*, 2023, **35**, 2302146.